\documentclass{ws-procs11x85}

\usepackage{balance}

\def\Journal#1#2#3#4{{#1} {\bf #2}, #3 (#4)}

\makeindex

\newcommand{\met} {\hbox{E\kern-0.5em\lower-0.1ex\hbox{/}}_T}

\begin{document}

\title{Top Quark Measurements at the Fermilab Tevatron}

\author{P. Azzi}

\address{I.N.F.N.-Sezione di Padova, Via Marzolo 8,
35100 Padova, Italy\\E-mail: azzi@pd.infn.it}


\twocolumn[\maketitle\abstract{
The top quark, discovered at the Tevatron in 1995, is a very interesting particle. Precise measurement of the top properties using large data samples will allow stringent tests of the Standard Model and offer a unique window on new physics. 
This report contains a review of the status of the current knowledge of the top quark as provided by the Run I results of the CDF and D0 experiments. A first look at various preliminary measurements obtained with data collected during Run II will also be presented.}]

\baselineskip=13.07pt
\section{Introduction}
The top quark is one of the building blocks of the Standard Model of Electroweak interactions as the partner of the bottom quark in the SU(2) isospin doublet of the third family of quarks.
After having eluded experimentalists for many years, the CDF experiment found evidence for its existence in 1994\cite{evidenceCDF} followed soon by its discovery by the CDF\cite{observationCDF} and D0\cite{observationD0} experiments in 1995. 
\par
While the data set collected at the Tevatron $p\bar p$ collider during Run I (${\cal{L}}_{int}=110$\,pb$^{-1}$) has been sufficient for the discovery of the top quark, it is still too small to allow stringent tests of the Standard Model in top enriched samples. The ensemble of available top measurements up-to-date is thus consistent with expectations, however Run I left us a few interesting discrepancies that will need to be addressed in Run II.
The large Run II dataset will allow one to achieve a greater precision and probe possible deviations from the Standard Model in a significant way.

\section{Top Production and Decay}
At the Tevatron the top quark is produced mainly in pairs through the process $qq,gg\rightarrow t\bar t$ with a cross section of about $6.7$\,pb\cite{kidonakis} for $m_t=175$\,GeV/c$^2$ and $\sqrt{s}=1.96$\,TeV. 
Within the Standard Model framework the top decays almost 100\% of the time via $t\rightarrow Wb$. Therefore it is customary to classify the $t\bar t$ final states based on the $W$ decays modes: dileptons ($\ell = e,\mu$), lepton+jets ($\ell = e,\mu$), all-jets and inclusive $\tau$ (hadronically decaying) final-state events.
\par
Even if its production cross section is much smaller compared to other Standard Model background processes, top events have very distinctive signatures that guide the overall analysis strategy: given the large top mass the decay products (leptons, jets) have large $p_T$'s and the event topology is central and spherical, and there are always two $b$ jets in the final state. Excellent lepton identification, energy resolution and $b$ identification capabilities are essential for a successful Run II top physics program. The full kinematical and heavy flavor characterization of top enriched data sets in terms of all the known Standard Model processes is important not only for precision measurements but also to test for any new phenomena. 

\section{The Run II Tevatron Collider and Detector Upgrades}
For Run II the Fermilab accelerator complex underwent a major upgrade. As a result the Tevatron operates at a higher center-of-mass energy $\sqrt{s}=1.96$\,TeV, with a bunch crossing interval of $396$\,ns with a goal of ${\cal{L}}_{inst}=33\times 10^{31}$\,cm$^{-2}$\,s$^{-1}$. By August 2003, the total integrated luminosity delivered by the Tevatron amounted to about  ${\cal{L}}_{int}=300$\,pb$^{-1}$ with a typical  ${\cal{L}}_{inst}=3$\,--\,$4.5\times 10^{31}$\,cm$^{-2}$\,s$^{-1}$. 
\par
The CDF and D0 detectors underwent extensive upgrades for Run II motivated by the need to improve overall acceptance, secondary vertex capabilities and to cope with the upgraded Tevatron performance.
CDF has retained its central calorimeter and part of the muon detectors, while it has replaced the central drift chamber (COT) and the silicon tracking system (L00, SVXII, ISL). New plug calorimeters and additional muon coverage allow CDF to extend lepton identification in the forward region. The most important upgrade of the D0 detector is the new tracking system which consists of a fiber tracker plus a silicon tracker immersed in a new $2$\,T superconducting solenoid. D0 has also improved the muon coverage and added new preshower detectors. Both CDF and D0 have new DAQ and trigger systems to cope with the shorter interbunch time. 

\subsection{Silicon Detectors Upgrades and High {\boldmath $p_T$ $b$-tagging}} 

The feasibility and the techniques of $b$ hadron identification at hadron machines have been firmly established in Run I. The use of $b$-tagging has been crucial for top discovery and it is an essential piece of the Run II top and exotic physics program. 
Both CDF and D0 silicon detectors upgrades followed the same guidelines: provide good coverage of the long ($\sim 30$\,cm) Tevatron luminous region, extend acceptance in the forward region, and with 3D reconstruction capabilities plus excellent impact parameter resolution achieve large signal efficiency and good background rejection.
\par 
The ``silicon vertex'' methods for identifying a $b$ jet in a top event exploit the $b$ hadron's long lifetime, large boost and significant charged track multiplicity of the decay. However, alternative methods (``soft lepton'') that exploit the softer $p_T$ spectrum and low isolation properties of $b$ semileptonic decay modes are also employed successfully by CDF and D0. 

\section{Top Cross Section Measurement}

Precise measurements of the $t\bar t$ production cross section and the branching ratios in all the decay channels provide a stringent test for the presence of new physics phenomena. Top-color and SUSY models predict not only alternative top production processes but extra decay modes that can alter the branching ratios of the various channels. The Run I top cross section measurements are summarized in Table \ref{tab:runIxs}.
The goal for Run II (${\cal{L}}_{int}=2$\,fb$^{-1}$) is to achieve a relative uncertainty of about $10\%$ or less on $\sigma_{t\bar t}$, this will be possible not only through the increased detector acceptance and efficiencies but also because the main data driven systematic uncertainties (jet energy scale, ISR/FSR, $\epsilon_{btag}$) will scale with the size of the control sample used for their determination. 

\begin{table*}[t]
\caption{Summary table of Run I $\sigma_{t\bar t}$ measurements, ${\cal {L}}_{int}=110$\,pb$^{-1}$.\label{tab:runIxs}}
\vspace{0.2cm}
\begin{center}
\tablefont{
\begin{tabular}{|c|c|c|}
\hline
\raisebox{0pt}[13pt][7pt]{$\sigma_{t\bar t}$(pb)} Channel &
\raisebox{0pt}[13pt][7pt]{CDF measurement} &
\raisebox{0pt}[13pt][7pt]{D0 measurement} \\
\hline
Dilepton & $8.4^{+4.5}_{-3.5}$ & $6.4\pm 3.4$ \\
Lepton+jets & $5.7^{+1.9}_{-1.5}$ & $5.2\pm 1.8$ \\
All jets & $7.6^{+3.5}_{-2.7}$ & $7.1\pm 3.2$ \\
Combined & $6.5^{+1.7}_{-1.4}$ & $5.9\pm 1.7$ \\
\hline
\end{tabular}
}
\end{center}
\end{table*}

\subsection{Run II Cross Section Measurement}

\begin{figure}
\psfig{figure=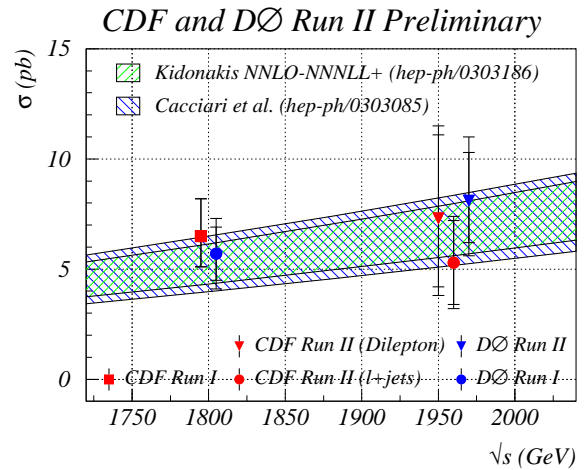,width=8.0truecm}
\caption{Summary of $\sigma_{t\bar t}$ versus center-of-mass energy.
\label{fig:xs_vs_cm}\vspace*{12pt}}
\end{figure}

The first Run II measurements of $\sigma_{t\bar t}$ have been focused on the channels with the highest signal-to-background ratio, namely the dilepton and lepton plus jets ($\ell$=$e$, $\mu$). 
The results presented have been summarized in Table \ref{tab:runIIxs} and they are shown in Fig.~\ref{fig:xs_vs_cm} as a function of the center-of-mass energy.

\begin{table*}[t]
\caption{Summary table of Run II $\sigma_{t\bar t}$ measurements.\label{tab:runIIxs} CDF and D0 preliminary results.}
\vspace{0.2cm}
\begin{center}
\tablefont{
\begin{tabular}{|c|c|c|c|c|}
\hline
\raisebox{0pt}[13pt][7pt]{Decay Channel} &
\raisebox{0pt}[13pt][7pt]{Method} &
\raisebox{0pt}[13pt][7pt]{$\sigma_{t\bar t}$(pb)} &
\raisebox{0pt}[13pt][7pt]{${\cal{L}}_{int}$ (pb$^{-1}$)} & 
\raisebox{0pt}[13pt][7pt]{Experiment} \\
\hline
Dilepton  & $\ell\ell$ & $8.7^{+6.4}_{-4.7}(stat)+^{+2.7}_{-2.0}(syst)\pm0.9(lum)$  & $90$\,--\,$107$ & D0 \\
Dilepton  & $\ell\ell$ & 7.6$^{+3.8}_{-3.1}(stat)^{+1.5}_{-1.9}(syst)$              & $126$ & CDF \\
Dilepton  & $\ell$+track & $7.3\pm 3.4(stat)\pm 1.7(syst)$                          & $126$ & CDF \\
\hline
$\ell$+jets & CSIP & $7.4^{+4.4}_{-3.6}(stat)^{+2.1}_{-1.6}(syst)\pm 0.7(lum)$      & $45$ & D0\\
$\ell$+jets & SVT &  $10.8^{+4.9}_{-4.0}(stat)^{+2.1}_{-2.0}(syst)\pm 1.1(lum)$     & $45$ & D0\\
$\ell$+jets & topo &  $4.6^{+3.1}_{-2.7}(stat)^{+2.1}_{-2.0}(syst)\pm 0.5(lum)$     & $92$ & D0\\
$\ell$+jets & SMT  &  $11.4^{+4.1}_{-3.5}(stat)^{+2.0}_{-1.8}(syst)\pm 1.1(lum)$    & $92$ & D0\\
$\ell$+jets & combined &  $8.0^{+2.4}_{-2.1}(stat)^{+1.7}_{-1.5}(syst)\pm 0.8(lum)$ & $92$ & D0\\
$\ell$+jets & SVX  & $5.3\pm 1.9(stat)\pm 0.9(syst)$                                & $57$ & CDF\\
$\ell$+jets & $H_T$ & $5.1\pm 1.8(stat)\pm 2.1(syst)$                               & $126$ & CDF\\
\hline
Dilepton,$\ell$+jets  & combined & $8.1^{+2.2}_{-2.0}(stat)^{+1.6}_{-1.4}(syst)\pm 0.8(lum)$ & $90$\,--\,$107$ & D0\\
\hline
\end{tabular}
}
\end{center}
\end{table*}

\subsection{Dilepton Channel}
\begin{figure}
\center
\psfig{figure=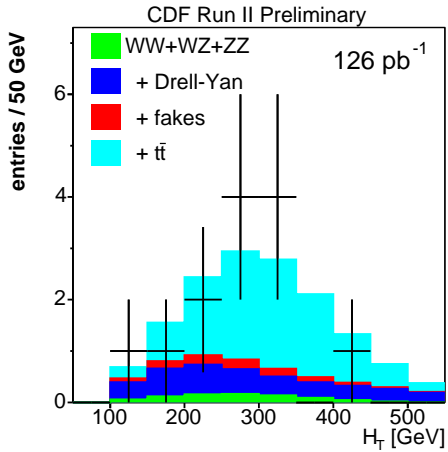,width=6.0truecm}
\caption{Run II data $H_T$ distribution of dilepton events compared to SM expectation (CDF).}
\label{fig:ht_dil}
\end{figure}

\begin{figure}
\center
\psfig{figure=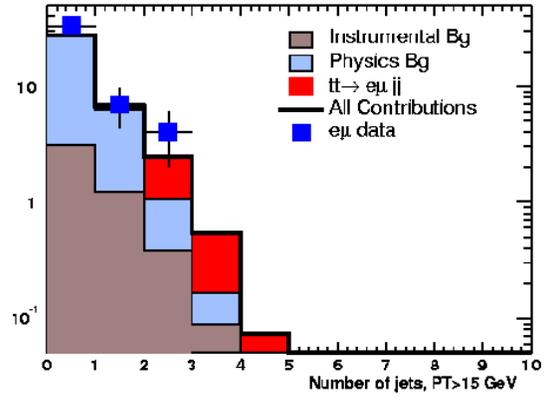,width=8.0truecm}
\caption{Run II data jet multiplicity distribution for dilepton events compared to SM expectation (D0).}
\label{fig:njet_dil}
\end{figure}

This final state is characterized by the presence of two high $p_T$ leptons (a reconstructed $e,\mu,\tau$ or an isolated track), large missing transverse energy from the missing neutrinos and two or more central jets. 
The main sources of background for this channel come from other Standard Model processes with similar signatures (Drell-Yan $\gamma^*/Z^0\rightarrow e^+e^-,\mu ^+\mu^-$, $Z^0\rightarrow \tau\tau$, $W^+W^-/W^{\pm}Z^0$) and processes with one real lepton and another object that fakes the second lepton. 
The dilepton event selection starts with two oppositely charged high $p_T$ leptons, asking that one or both are well isolated from nearby track activity. Different techniques are employed in order to reduce the contribution from $Z^0$ events without reducing the signal acceptance. $\met$ is required to be large given the two neutrinos from $W$ decay and the presence of two central jets accounts for the two $b$'s from the top decay. Other kinematical and topological cuts ($H_T$, total energy of all the object in the event, $\Delta\phi(\met,object)$) are finally employed in order to reduce the remaining backgrounds. The comparison of the remaining events after selection with the total Standard Model expectation (background plus signal) is shown in Fig.~\ref{fig:ht_dil} and Fig.~\ref{fig:njet_dil}.   

\subsection{Lepton Plus Jets Channel}
\begin{figure}
\center
\psfig{figure=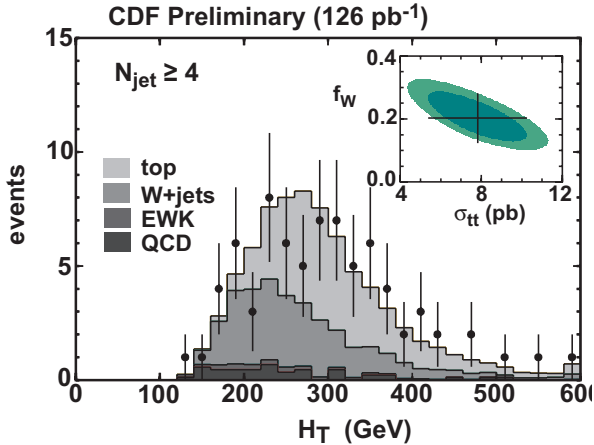,width=8.0truecm}
\caption{Run II data $H_T$ distribution for lepton plus jets events with at least four jets compared to SM 
expectations (CDF).}
\label{fig:CDFljets_kin}
\end{figure}

\begin{figure}
\center
\psfig{figure=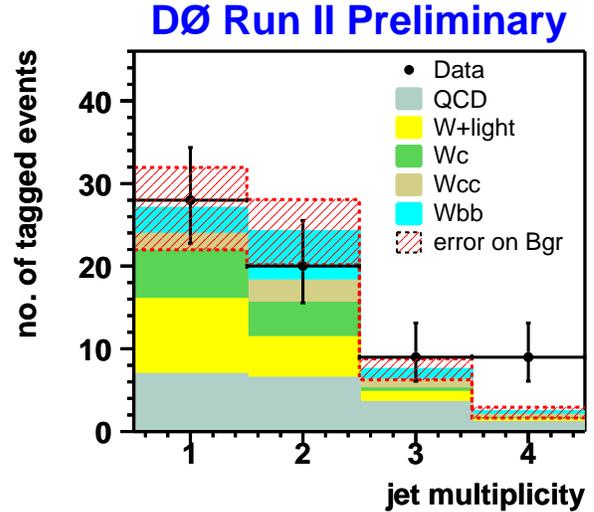,width=8.0truecm}
\caption{Run II data jet multiplicity distribution in the lepton plus jets channel with a $b$-tagged jet compared with SM expectations (D0).}
\label{fig:njet_ljettag}
\end{figure}

The lepton plus jets signature is characterized by the presence of one high $p_T$ lepton and large $\met$ due to the leptonic $W$ decay plus three or more jets from the hadronically decaying $W$ and the $b$ jets. 
This final state suffers from large $W+jets$ background. However, kinematical and topological properties of the $t\bar t$ signal or its heavy flavor content (or both) provide a good separation from the background processes. 
In the kinematical approach, after the basic event selection, variables such as $H_T$, the scalar sum of all the objects' transverse energies in the event, or the aplanarity ${\cal{A}}$, a measure of the event shape, are found to be the most discriminant, see for example Fig.~\ref{fig:CDFljets_kin}. 
A complementary approach is to exploit the heavy flavor content of signal events and the large $b$-tagging efficiency compared to the low fake rate. There are several $b$ identification algorithms available at the moment, some employ the silicon vertex detector information, while another category focuses on the peculiar properties of leptons from $b$ semileptonic decays. In Fig.~\ref{fig:njet_ljettag} the jet multiplicity in $W+jets$ events is shown after the requirement of an identified secondary vertex: the points are the data compared to the background Standard Model expectation. The excess due to the $t\bar t$ signal is visible in the three or more jet bins. 

\subsection{All Hadronic Channel}
The all-jets final state where both $W$ decay hadronically is a very challenging signature of six central and energetic jets, swamped by a QCD multijet background of several orders of magnitude bigger than the $t\bar t$ signal. However, in Run I both experiments succeeded in isolating the signal and measuring cross section and mass in this channel as well\rlap.\,\cite{CDFtophad}$^,$\cite{D0tophad} The D0 experiment has taken a first look at this channel in Run II repeating the Run I neural network analysis. A small excess of $78$ events with a background of $68\pm 1.6$ is found in ${\cal{L}}_{int}=80.7$\,pb$^{-1}$ of data, consistent with Standard Model expectations. The measurement of the cross section in this channel is in progress. 

\subsection{Test for New Physics in {\boldmath $t\bar t$} Production}

\begin{figure}
\center
\psfig{figure=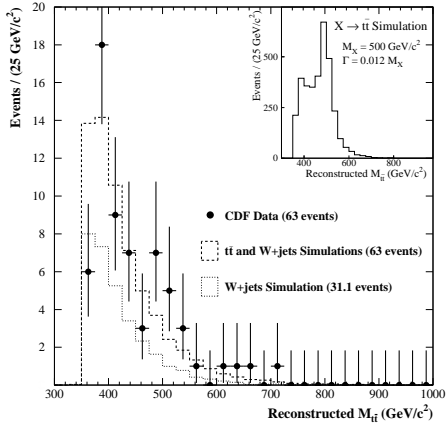,width=6.0truecm}
\caption{Run I data $M(t\bar t)$ distribution (CDF).}
\label{fig:CDF_mtt}
\end{figure}

\begin{figure}
\center
\psfig{figure=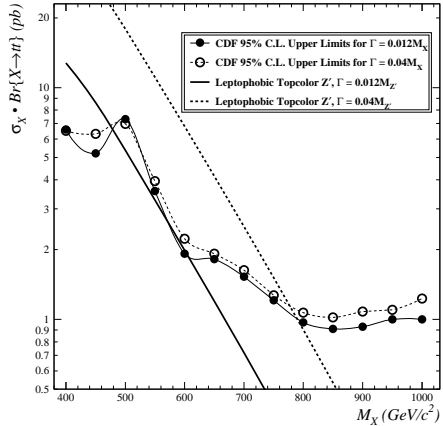,width=6.0truecm}
\caption{Run I limit on a leptophobic $Z^\prime\rightarrow t\bar t$ (CDF).}
\label{fig:CDF_lim}
\end{figure}

Both CDF\cite{CDF_mtt} and D0\cite{D0_mtt} have searched for $t\bar t$ resonances using the Run I data sample. Models with a dynamically broken EW symmetry (technicolor) predict a top-quark condensate, $X$, that decays to a $t\bar t$ pair. By searching for narrow $t\bar t$ resonances this limit becomes model independent, see Fig.~\ref{fig:CDF_mtt}. However, 95\% CL limits have been placed on a leptophobic $Z^\prime\rightarrow t\bar t$ with a large cross section for $m(Z^\prime)<560$\,GeV/c$^2$ by both CDF and D0, see Fig.~\ref{fig:CDF_lim}.

\begin{figure}
\center
\psfig{figure=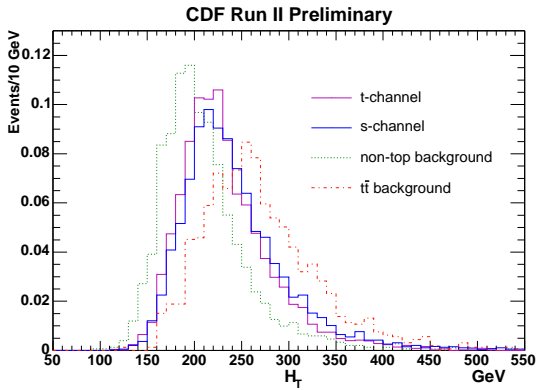,width=8.0truecm}
\caption{$H_T$ distribution for MC single top events compared to $t\bar t$ and $W+jets$ (CDF).}
\label{fig:CDF_singletopHt}
\end{figure}
\begin{figure}
\center
\psfig{figure=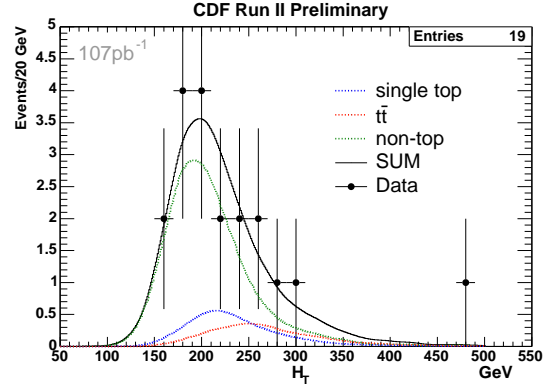,width=8.0truecm}
\caption{Run II data (points) $H_T$ distribution compared to the expected  SM background plus $t\bar t$ and single top production (CDF).}
\label{fig:CDF_singletop}
\end{figure}
\section{Single Top Physics}
In addition to pair production, single top quarks can be produced by weak interaction by a virtual $W$ or through $Wg$ fusion, with a total cross section of about $\sigma_{tX}= 2.9$\,pb\rlap.\,\cite{singletop} Single top production is interesting in its own right: a precise measurement of the cross section would provide a direct determination of $|V_{tb}|$ with a $14\%$ uncertainty expected for ${\cal{L}}_{int}=2$\,fb$^{-1}$ of data. Moreover single top events have the same final-state experimental signature as the Standard Model Higgs associated production process ($HW\rightarrow b\bar b \ell\nu_{\ell}$). The extraction of a single top signal is more challenging than the pair produced case since there are fewer objects in the final state and the overall event properties are less distinct from the $W+jets$ background, see Fig.~\ref{fig:CDF_singletopHt}. 
In Run I searches for single top production (in the $s$- and $t$-channels separately and combined) were performed by both CDF\cite{CDFsingletop} and D0\rlap.\,\cite{D0singletop} The same search method has been applied by CDF on a Run II dataset of about  ${\cal{L}}_{int}=107$\,pb$^{-1}$ and the preliminary result is still consistent with the Run I cross section limit, $\sigma_{tX}^{RunII}(comb)<17.5$\,pb\,@~95\%CL. The $H_T$ distribution for the candidate events compared to Standard Model expectation for signal and background is shown in Fig.~\ref{fig:CDF_singletop}.

\section{$W$ Helicity in Top Decays}
\begin{figure}
\center
\psfig{figure=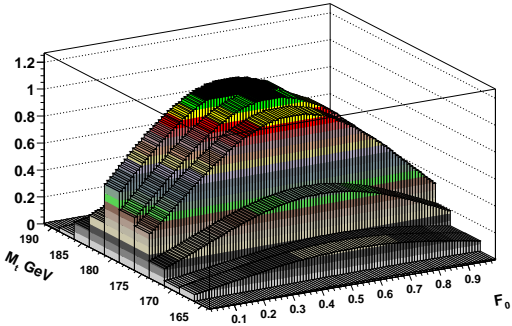,width=8.0truecm}
\caption{Two dimensional ($m_t$,$f_0$) probability distribution of Run I lepton plus jets data (D0).}
\label{fig:D0Whel}
\end{figure}

Since the top is the only quark that decays free without hadronizing, the decay products carry its polarization information. In the Standard Model the top quark decays only to longitudinal or left-handed $W$'s, where the ratio is predicted to be about $f_0=\frac{W_{long}}{W_{left}}=70\%$ in the case of a $m_t=175$\,GeV/c$^2$. The helicity information is reflected in several kinematical properties of the decay products ($W$ lepton $p_T$, $M(\ell b)$) that are traditionally used to extract an experimental measurement of $f_0$. However, D0 has performed a new measurement using the Run I datasets: $f_0= 0.56\pm 0.31(stat)\pm 0.04(syst)$ under the assumption of $m_t=175$\,GeV/c$^2$, see in Fig.~\ref{fig:D0Whel} the two dimensional ($m_t$,$f_0$) probability distribution. The likelihood method used here makes better use of the event information thus greatly improving the statistical uncertainty. This method is used also in the measurement of the top mass and will be discussed more in Sec.~\ref{sec:topmass}. 

\section{Top Mass}\label{sec:topmass}
\begin{figure}
\center
\psfig{figure=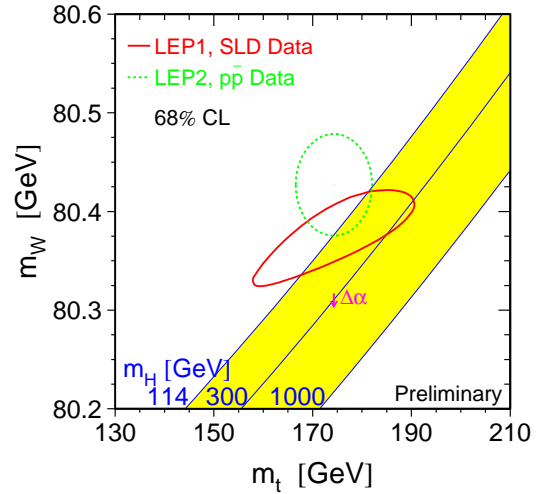,width=7.0truecm}
\caption{The closed curves representing experimental measurements of $m_t$ and $m_W$ constrain the SM Higgs mass. The shaded band shows the allowed combinations of $m_t$ and $m_W$ for various $m_H$.}
\label{fig:MtMw}
\end{figure}

The top quark mass is a fundamental Standard Model parameter that needs to be measured with the greatest possible precision. It is needed to determine the strength of the $ttH$ coupling and it has a substantial effect on radiative corrections. In fact, an uncertainty of $2$\,GeV/c$^2$ on the top mass would constrain the Higgs mass to $35\%$, see Fig.~\ref{fig:MtMw}.
It is not an easy task to achieve such a small uncertainty, but several experimental handles are available in Run II. On one hand the increased detector acceptance and large data sample will allow one to select purer samples less sensitive to systematic uncertainties: for instance requiring events with well measured jets (lowers the energy scale uncertainty) and two $b$-tagged jets (lowers the overall background). On the other hand, since most of the systematics are data driven, their uncertainty will scale approximately with $1/\sqrt{N}$, $N$ being the number of events of the control samples themselves.    

\begin{figure*}[t]
\psfig{figure=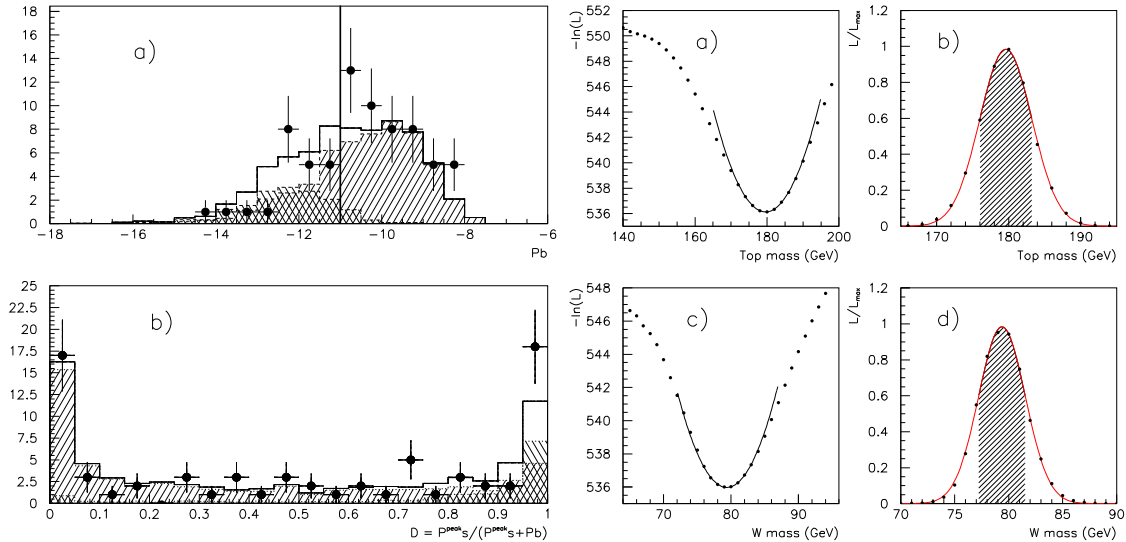,width=15truecm}
\caption{New D0 Run I top mass measurement. Left: a) background probability distribution; b) ratio $P_{top}/(P_{bkg}+P_{top})$. Right: a) and b) fitted top mass value and its uncertainty; c) and d) fitted $W$ mass when the top mass is fixed to its fitted value.}
\label{fig:D0mass} 
\end{figure*}

\subsection{Run I Measurement Summary}

Using the Run I dataset the CDF and D0 experiment have measured directly the top mass in channels (lepton+jets, dilepton and all hadronic) employing different methods and techniques. The results from the two experiments and their combination are summarized in Table~\ref{tab:runImass}. However, the single most precise measurement on the Run I data comes from the latest measurement from the D0 experiment in the lepton plus jets channel\cite{newD0mass} of $m_t=180.1\pm 5.4$\,GeV/c$^2$. The likelihood method employed for this measurement was originally proposed for the mass reconstruction in dilepton events\cite{dalitz}$^-$\cite{D0dilmass} where the system is underconstrained for a simple kinematic fit. However, the technique is very useful also in the case of lepton plus jets events, since a better use of the event information effectively increases the statistical power of the data sample itself. Each event has an associated probability to be signal or background defined in terms of the matrix element information, and this probability is convolved with a transfer function that relates the object at the parton level to the object after reconstruction. The only background considered is $W+ 4\,jets$, which makes up 80\%\ of the total, and cleanup cuts are added to the data to further reduce the absolute background contribution. The remaining 22 events in ${\cal{L}}=125$\,pb$^{-1}$ of data are then used in a global likelihood fit to extract the top mass and the $W$ mass (or the $f_0$ at the same time), as shown in Fig.~\ref{fig:D0mass}. The large reduction in statistical uncertainty, corresponding to an effective increase of the data set by a factor 2.4, is mainly due to the more complete use of the event information. 

\begin{table*}[t]
\caption{Summary table of Run I $m_t$ measurements.\label{tab:runImass}}
\vspace{0.2cm}
\begin{center}
\tablefont{
\begin{tabular}{|c|c|c|}
\hline
\raisebox{0pt}[13pt][7pt]{$m_t$ (GeV/c$^2$)} Channel &
\raisebox{0pt}[13pt][7pt]{CDF measurement} &
\raisebox{0pt}[13pt][7pt]{D0 measurement} \\
\hline
Dilepton & $167.4\pm 11.4$ & $168.4\pm 12.8$ \\
Lepton+jets & $175.9\pm 7.1$ & $173.3\pm 7.8$ \\
All jets & $186.0\pm 11.5$ & --  \\
Combined & $176.0\pm 6.5$ & $172.1\pm 7.1$ \\
\hline
CDF+D0 Combined & \multicolumn{2}{|c|}{$174.3\pm 5.1$}\\
\hline
Lepton+jets (New) &  -- & $180.1\pm 5.4$\\
\hline
\end{tabular}
}
\end{center}
\end{table*}

\subsection{Issues for Precision Top Mass in Run II}
A precise measurement of the top mass combines cutting edge theoretical knowledge with state-of-the-art detector calibration. The highest contribution to the systematic uncertainty still comes from the jet energy scale. 
With the statistics available now the best calibration sample consists of events where a jet is recoiling against a well measured photon, with larger statistics the sample where a jet recoils against a reconstructed $Z$ can be used. Finally the hadronic $W$ in lepton plus jets events can provide an in situ calibration for the light quark jets, while the $Z\rightarrow b\bar b$ signal would be used for the heavy flavor ones.
A large amount of data will allow one to not only reduce the systematics above but also to pick the best measured event categories with smaller backgrounds and that are less sensitive to systematics uncertainties. 
However, to achieve the ultimate precision excellent Monte Carlo generators implementing the latest theory knowledge and understanding of all the various effects (ISR, FSR, PDF's) plus an accurate detector simulation are essential. 
\par
\begin{figure}
\center
\psfig{figure=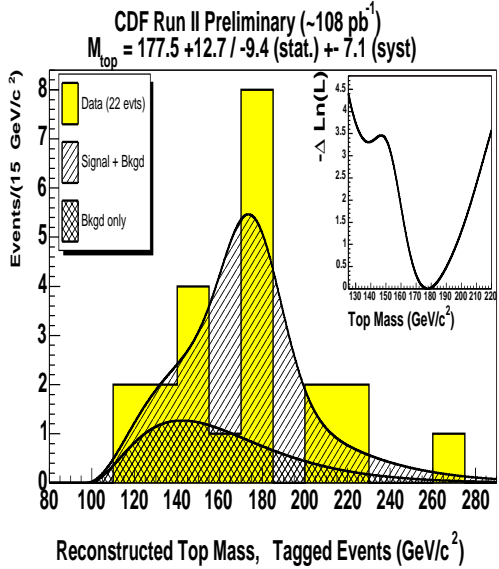,width=7.0truecm,height=7.0truecm}
\caption{Reconstructed $m_t$ distribution in the lepton+jets channel with a $b$-tagged jet in Run II data (CDF).}
\label{fig:CDFmass_ljet}
\end{figure}

\begin{figure}
\center
\psfig{figure=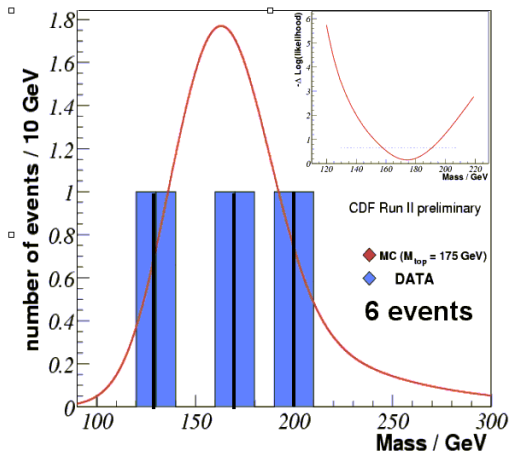,width=7.0truecm}
\caption{Reconstructed $m_t$ in the dilepton channel in Run II data (CDF).}
\label{fig:CDFmass_dil}
\end{figure}

 
While work is in progress on all these fronts, preliminary measurements of the top mass, still dominated by large systematic uncertainties, have been performed using the Run II data sample.
In the lepton plus four jets channel with at least one secondary vertex $b$-tagged jet a value of $m_t = 177.5^{+12.7}_{-9.4}(stat)\pm 7.1(syst)$\,GeV/c$^2$ is found using 22 candidate events shown in Fig.~\ref{fig:CDFmass_ljet}. In the dilepton channel a preliminary measurement of 
$m_t=175.0^{+17.4}_{-16.9}(stat)\pm 7.9(syst)$\,GeV/c$^2$, is obtained using 6 candidate events, shown in Fig.~\ref{fig:CDFmass_dil}.

\section{Conclusions}

The top is still a very young particle and our current knowledge about its properties comes from the Run I Tevatron measurements. This accelerator and its two experiments, CDF and D0, are the place for top physics still for years to come. In Run II a data sample about 50 times the Run I statistics will be collected. It will be possible to achieve better precision in the measurements and perform significant tests of the Standard Model expectations. Maybe there will be surprises ahead... The first preliminary round of Run II measurements of the production cross section and mass covers already a variety of channels and very soon the uncertainties will start to drop. 
The two experiments are exploiting all their new upgraded detector features and a very exciting top physics program lies ahead.

\section*{Acknowledgments}
I'd like to thank Keith Ellis and the LP2003 conference organizers for this great opportunity. I'd like also to thank my friends and colleagues Marumi Kado, Ela Barberis, Aurelio Juste, Jaco Konigsberg and Simona Rolli for their help in the preparation of this report.

\clearpage
\balance
\twocolumn[
\section*{DISCUSSION}
]

\begin{description}
\item[S. Heinemeyer] (LMU Munich):
What is the new combined value for $m_t$ from CDF Run I plus Run II plus new D0 Run I result? 

\item[Patrizia Azzi{\rm :}]
The combined number is not available yet.

\item[R. Keeler] (U. Victoria):
Was the choice of the ``future position'' of the result from the collider on the $m_t$ versus $m_W$ plane motivated by any experimental knowledge? 

\item[Patrizia Azzi{\rm :}]
No. It was for illustration only.

\end{description}


\begin{thebibliography}{9}

\bibitem{evidenceCDF} F.Abe {\it et al.}, \Journal{{\em Phys. Rev. Lett.}}{73}{225}{1994}.
\bibitem{observationCDF} F.Abe {\it et al.}, \Journal{{\em Phys. Rev. Lett}}{74}{2626}{1995}.
\bibitem{observationD0} S.Abachi {\it et al.}, \Journal{{\em Phys. Rev. Lett.} }{74}{2632}{1995}.
\bibitem{kidonakis} M. Cacciari {\it et al.}, {\em{hep-ph/0303085}}, \\
 N.Kidonakis and  R.Vogt, {\em{hep-ph/0308222}}.
\bibitem{CDFtophad} F.Abe {\it et al.}, \Journal{{\em Phys. Rev. Lett.} }{79}{1992}{1997}.
\bibitem{D0tophad} D0 Collaboration, \Journal{{\em Phys. Rev.} D}{60}{012001}{1999}.
\bibitem{CDF_mtt} T.Affolder {\it et al.}, \Journal{{\em Phys. Rev. Lett.}}{85}{2062}{2000}.
\bibitem{D0_mtt} V.M.Abazov {\it et al.}, {\em {hep-ex/0307079}}{2003}.
\bibitem{singletop} B.W.Harris {\it et al.}, \Journal{{\em Phys. Rev.} D}{66}{054024}{2002}.
\bibitem{CDFsingletop} D.Acosta {\it et al.}, \Journal{{\em Phys. Rev.} D}{65}{091102}{2002}.
\bibitem{D0singletop} V.Abazov {\it et al.}, \Journal{{\em Phys. Lett.} B}{517}{282}{2001}.
\bibitem{newD0mass} J. Estrada {\em for D0 Coll.}, {\em hep-ex/03/02031}{2002}.
\bibitem{dalitz} R.H.Dalitz and G.R.Goldstein, \Journal{{\em Proc. R. Soc. Lond.}}{A445}{2803}{1999}.
\bibitem{kondo} K. Kondo {\em et al.}, \Journal{\em J.Phys.Soc.Jap}{62}{1177}{1993}.
\bibitem{D0dilmass} B. Abbott {\em et al.}, \Journal{{\em Phys. Rev} D}{60}{052001}{1999}.



\end{thebibliography}
\end{document}